\RequirePackage{fix-cm}
\documentclass{svjour3}                     
\smartqed  
\usepackage{graphicx}
\usepackage{amsfonts}

 \journalname{Few-Body Systems}
\begin{document}

\title{Fine-tuning of the quasi-bound $K^- pp$ state
\thanks{The work was supported by the THEIA, STRONG-2020 Project funded by the EU Framework Programme for Research and Innovation, Horizon 2020}
}

\titlerunning{Fine-tuning of the quasi-bound $K^- pp$ state}

\author{N.V. Shevchenko 
}


\institute{N.V. Shevchenko \at
              Nuclear Physics Institute, 25068 \v{R}e\v{z}, Czech Republic \\
              \email{shevchenko@ujf.cas.cz} 
}

\date{Received: date / Accepted: date}

\maketitle

\begin{abstract}

Characteristics of the quasi-bound state in the $K^- pp$ system strongly depend on the model of
antikaon-nucleon interaction and weakly - on the nucleon-nucleon potential. In the present paper,
dynamically exact Faddeev-type calculations with coupled $\bar{K}NN$ and $\pi \Sigma N$ were
performed using different models of the $\Sigma N$ and $\pi N$ interactions to study the
influence of these "less important" interaction on the three-body result. In addition,
dynamically exact three-body Faddeev-type AGS calculations with three coupled particle channels
$\bar{K}NN - \pi \Sigma N - \pi \Lambda N$ were performed.

\end{abstract}


\section{Introduction}
\label{intro_sect}

The attractive nature of $\bar{K}N$ interaction has stimulated theoretical and experimental searches
for $K^-$ bound states in different systems. In particular, many theoretical calculations devoted to
the lightest possible system $\bar{K}NN$ were performed using different methods and inputs. All of them
agree that a quasi-bound state in the $K^- pp$ system exists, but they yield quite diverse binding energies
and widths. The experimental situation is also unsettled: several evidences for the $K^- pp$ state were
reported by several experimental groups, but the estimated binding energies and especially decay widths
of such states differ from each other and are far from all theoretical predictions. The most recent
E15 experiment at J-PARC \cite{E15_JPARC} reported the first clear signal of the $\bar{K}NN$ quasi-bound
state with binding energy $42 \pm 3 (\rm stat.) {}^{+3}_{-4} (\rm syst.)$ MeV, and width
$100 \pm 7 (\rm stat.) {}^{+19}_{-9} (\rm syst.)$ MeV. The measured binding energy of the state
is comparable with some of the theoretical results, while the width is much larger than the theoretical predictions.

We studied the quasi-bound state in the $K^- pp$ system solving Faddeev-type dynamically exact three-body 
Alt-Grassberger-Sandhas (AGS) equations \cite{AGS3} with coupled $\bar{K}NN - \pi \Sigma N$ channels with
different input. The most recent results  of our calculations can be found in \cite{Kd_Kpp_last}. 
In particular, three models of the $\bar{K}N$ interaction were used, and calculations with two of them
give $K^- pp$ binding energies close to the experimental values \cite{E15_JPARC}. However, the experimental
width measured by the E15 experiment at J-PARC is much larger than our predictions. In our previous calculations
we demonstrated that  the antikaon-nucleon interaction plays the main role in the description of the
$K^- pp$ system. The binding energy and width of the quasi-bound state in the system strongly depend on it. 
On the contrary, dependence on the nucleon-nucleon potential is weak.  The present calculations aim to check,
what else could change the theoretical predictions for the characteristics of the $K^- pp$ quasi-bound state.
To do this we studied the dependence of the binding energy and width of the quasi-bound state 
on the $\Sigma N - \Lambda N$ interaction models and used different versions of the $\pi N$ potential,
excluded from the previous calculations. In addition, we performed for the first time dynamically exact
three-body Faddeev-type AGS calculations with three coupled channels $\bar{K}NN$, $\pi \Sigma N$, and
$\pi \Lambda N$.  For this sake new $\bar{K}N - \pi \Sigma - \pi \Lambda$ potentials were constructed
and used. 

The paper is organized as follows. The three-body AGS equations with two coupled $\bar{K}NN - \pi \Sigma N$
channels are described in Section \ref{3AGS2ch_sect}, the section after it is devoted to the two-body 
interaction models used in the calculations. Section \ref{3AGS3ch_sect} contains description of the
AGS equations with three coupled $\bar{K}NN - \pi \Sigma N - \pi \Lambda N$ channels and the
corresponding $\bar{K}N - \pi \Sigma- \pi \Lambda$  antikaon-nucleon potentials. The results of the three-body
calculations are separated at those obtained with two coupled $\bar{K}NN - \pi \Sigma N$
(Section \ref{results2ch_sect}) and three coupled $\bar{K}NN - \pi \Sigma N - \pi \Lambda N$ channels
(Section \ref{results3ch_sect}). The paper is finished with the Summary.

\section{Three-body AGS equations with two coupled $\bar{K}NN - \pi \Sigma N$ channels}
\label{3AGS2ch_sect}

The quasi-bound state in the $K^- pp$ system was studied using dynamically exact Faddeev-type
AGS equations  \cite{AGS3}. The system of three-body equations was written for separable potentials
${V}_{i}^{\alpha \beta}$ leading to separable $T$-matrices ${T}_{i}^{\alpha \beta}$
\begin{equation}
\label{VTsep}
{V}_{i}^{\alpha \beta} =  \lambda_i^{\alpha \beta} | {g}_{i}^{\alpha} \rangle 
  \langle {g}_{i}^{\beta} | \qquad \to
\qquad
{T}_{i}^{\alpha \beta} =  | {g}_{i}^{\alpha} \rangle 
  {\tau}_{i}^{\alpha \beta}   \langle {g}_{i}^{\beta} |,
\end{equation}
were one-term separable potentials are assumed for simplicity. The AGS equations in the operator
form
\begin{equation}
\label{AGS2ch}
 {X}_{ij}^{\alpha \beta}(z) = \delta_{\alpha \beta} {Z}_{ij}^{\alpha} + 
 \sum_{k=1}^3 \sum_{\gamma=1}^3  {Z}_{ik}^{\alpha} \, {\tau}_{k}^{\alpha \gamma} \,
   {X}_{kj}^{\gamma \beta}
\end{equation}
contain three-body transition ${X}_{ij}^{\alpha \beta}$ and kernel ${Z}_{ij}^{\alpha}$ operators.
Faddeev indices $i,j,k = 1,2,3$ in Eq.(\ref{AGS2ch}) simultaneously denote a pair of particles and
the corresponding spectator (which is the third particle).  Since antikaon-nucleon interaction
is coupled to $\pi \Sigma$ channel through $\Lambda(1405)$ resonance, the corresponding
potentials with coupled $\bar{K}N$ and $\pi \Sigma$ channels were used. The second, $\pi \Sigma$
channel was included into the three-body equations Eq.(\ref{AGS2ch}) directly, due to this additional
indices $\alpha, \beta, \gamma$ denoting a particle channel
\begin{equation}
\alpha = 1: | \bar{K}_1N_2N_3 \rangle, \qquad \alpha = 2: | \pi_1 \Sigma_2 N_3 \rangle,
\qquad \alpha = 3: | \pi_1 N_2 \Sigma_3 \rangle
\end{equation}
were introduced.
The system of operator equations (\ref{AGS2ch}) was written down in momentum representation
and antisymmetrized due to identical fermions in the $\bar{K}NN$ channel. The corresponding system
of integral equations was solved numerically. The three-body system with spin $S^{(3)} = 0$ and 
isospin $I^{(3)}=1/2$ was studied.

\section{Two-body input}
\label{TwoBodyInput_sect}

The input for the three-body Faddeev-type AGS equations Eq.(\ref{AGS2ch}) are two-body
$T$-matrices Eq.(\ref{VTsep}), describing two-body interactions between all pairs of particles
in all particle channels. Since searching for quasibound states assumes working at low
energies, and low-energy interactions are satisfactorily described
by s waves, we used orbital momentum $l=0$ for all the two-body interactions.

\subsection{Antikaon-nucleon interaction}

Three our antikaon-nucleon interaction models, constructed before, were used in the present
calculations. Two of them are phenomenological potentials $V_{\bar{K}N}^{\rm 1,SIDD}$
and  $V_{\bar{K}N}^{\rm 2,SIDD}$ with one- or two-pole structure of the $\Lambda(1405)$
resonance correspondingly \cite{VKN_SIDD12}. The phenomenological potentials couple two channels:
$\bar{K}N$ and $\pi \Sigma$. The third, a chirally motivated potential
$V_{\bar{K}N}^{\rm Chiral}$ \cite{ourKNN_I} with two-pole structure of the $\Lambda(1405)$
resonance describes three coupled channels: $\bar{K}N$, $\pi \Sigma$, and $\pi \Lambda$.

All three potentials with the same level of accuracy reproduce low-energy experimental data
on $K^- p$ system. In particular, they reproduce $1s$ level shift and width of kaonic hydrogen 
caused by strong interaction and measured by SIDDHARTA collaboration \cite{SIDDHARTA}. 
The $1s$ level of the atom was calculated by us directly, without using some approximate
Deser-type formula: Lippmann-Schwinger equation for the strong $\bar{K}N-\pi \Sigma (- \pi \Lambda)$
plus Coulomb $K^- p$ potentials was solved. The experimental cross-sections of the elastic
$K^- p \to K^- p$ and inelastic $K^- p \to M B$ reactions 
\cite{Kp2exp,Kp3exp_1,Kp3exp_2,Kp4exp,Kp5exp,Kp6exp}
are also reproduced by all three of our potentials. Accurately measured threshold branching
ratios $\gamma$, $R_c$ and $R_n$ \cite{gammaKp1,gammaKp2} are reproduced by the chirally
motivated model $V_{\bar{K}N}^{\rm Chiral}$ directly. The phenomenological potentials
$V_{\bar{K}N}^{\rm 1,SIDD}$ and $V_{\bar{K}N}^{\rm 2,SIDD}$ reproduce experimental value of
$\gamma$ ratio directly, while the measured $R_c$ and $R_n$ were used for construction
of an arteficial $R_{\pi \Sigma}$ branchig ratio, reproduced by the potentials \cite{VKN_SIDD12}.
The reason is the absence of the directly included $\pi \Lambda$ channel in the phenomenological
antikaon-nucleon potentials.
Finally, all three $\bar{K}N$ interaction models reveal a bump in $\pi \Sigma$ elastic
cross-sections near the $\Lambda(1405)$ resonance mass
 $M_{\Lambda(1405)}^{PDG} = 1405.1^{+1.3}_{-1.0}$ MeV, according to the Particle Data Group 
\cite{PDG}.

\subsection{Nucleon-nucleon interaction}

Two-term Separable New potential (TSN) of nucleon-nucleon interaction was constructed
in \cite{Kd_Kpp_last}. It has a form
\begin{equation}
\label{VNN}
V_{NN}^{\rm TSN}(k,k') = \sum_{m=1}^2 g_{m}(k) \, \lambda_{m} \; g_{m}(k')
\end{equation}
with form-factors
\begin{equation}
\label{gNN}
g_{m}(k) = \sum_{n=1}^3 \frac{\gamma_{mn}}{(\beta_{mn})^2 + k^2}
\quad {\rm for \;} m=1,2.
\end{equation}
and parameters $\lambda_m$, $\gamma_{mn}$, $\beta_{mn}$
fitted to phase shifts of Argonne $V18$ potential \cite{ArgonneV18}.
The potential reproduces triplet and singlet scattering lengths $a$ and gives
effective ranges $r_{\rm eff}$
\begin{eqnarray}
a_{np}^{\rm TSN} = -5.400  {\, \rm fm}, \qquad  r_{{\rm eff}, np}^{\rm TSN} = 1.744 {\, \rm fm}, \\
a_{pp}^{\rm TSN} = 16.325 {\, \rm fm}, \qquad  r_{{\rm eff}, pp}^{\rm TSN} = 2.792 {\, \rm fm.}
\end{eqnarray}
Deuteron binding energy given by the $V_{NN}^{\rm TSN}$ is $E_{\rm{deu}} = 2.22$ MeV.

\subsection{$\Sigma N - \Lambda N$ interaction models}

A simple separable model of the $\Sigma N - \Lambda N$ potential was used in our previous
calculations \cite{Kd_Kpp_last}
\begin{equation}
\label{VSigN}
 V_{I,S}^{\Sigma N}(k,k') = \lambda_{I,S}^{\Sigma N} \;
g_{I,S}^{\Sigma N}(k) \, g_{I,S}^{\Sigma N}(k'), \quad
g_{I,S}^{\Sigma N} (k) = \frac{1}{(k^2 + \beta_{I,S}^{\Sigma N})^2}
\end{equation}
with parameters  $\lambda_{I,S}^{\Sigma N}$ and $\beta_{I,S}^{\Sigma N}$  fitted to not very
accurate experimental cross-sections  \cite{SigmaN1,SigmaN2,SigmaN3,SigmaN4,SigmaN5}.
The interaction strongly depends on the two-body isospin. 
The $\Sigma N$ interaction model in isospin $I=3/2$ state is a one-channel one, it has real
strength parameters $\lambda_{I,S}^{\Sigma N}$.  Isospin $I=1/2$ state $\Sigma N$ system is coupled
to the $\Lambda N$ channel, therefore a coupled-channel $\Sigma N - \Lambda N$ potential with real 
parameters $\lambda_{I,S}^{\Sigma N}$ was constructed. Since the lowest $\pi \Lambda N$ channel
is not directly included in the three-body calculations with two coupled $\bar{K}NN - \pi \Sigma N$
channels, the exact optical version of the coupled-channel potential $V_{I=1/2,S}^{\Sigma N}$ was used.
The strength parameters $\lambda_{I=1/2,S}^{\Sigma N}$ of the exact optical potential turn to energy
dependent functions. On the contrary, the three-body calculations with three coupled 
$\bar{K}NN - \pi \Sigma N - \pi \Lambda N$ channels were performed with the original version
of the $I=1/2$ potential with coupled $\Sigma N - \Lambda N$ channels.

New fits of the $\Sigma N - \Lambda N$ potential parameters Eq.(\ref{VSigN}) were performed
to  study dependence of the $K^- pp$ pole position on the $\Sigma N - \Lambda N$ interaction.
Two new sets are spin-dependent and spin-independent ones fitted to the experimental cross-sections
\cite{SigmaN1,SigmaN2,SigmaN3,SigmaN4,SigmaN5}. Two other sets were additionally fitted to
scattering lengths of the different charge states of $\Sigma N$ and $\Lambda N$ systems taken
from an "advanced" potential \cite{HyperN_poten}. All together four new sets of parameters
Eq.(\ref{VSigN}) were obtained for spin-dependent (denoted as "SDep")
or spin-independent ("SInd") potentials with or without fits to scattering length ("ScL" or "noScL").

\subsection{$\pi N$ potential}

It is known that $s$-wave $\pi N$ interaction is weak. Due to this the $\pi N$ potential $V^{\pi N}$
was neglected in our previous calculations of the $\bar{K}NN$ systems \cite{Kd_Kpp_last}. To check
dependence of the three-body results on the pion nucleon interaction, a one-term separable $\pi N$
potential was constructed
\begin{equation}
\label{VpiN}
 V_{I}^{\pi N}(k,k') = \lambda_{I}^{\pi N} \;
g_{I}^{\pi N}(k) \, g_{I}^{\pi N}(k'), \quad
g_{I}^{\pi N} (k) = \frac{1}{(k^2 + \beta_{I}^{\pi N})^2 }.
\end{equation}
First, four sets of  $ \lambda_{I}^{\pi N}$ and $ \beta_{I}^{\pi N}$ 
parameters (denoted as "1a", "1b", "2a", and "2b") were arbitrarily chosen with the only condition that
there are no bound states in the $\pi N$ system:
\begin{eqnarray}
&{}& \beta_{1/2}^{\pi N} = \beta_{3/2}^{\pi N} = 3 \; {\rm fm^{-1}}, \qquad
\lambda_{1/2}^{\pi N} = \lambda_{3/2}^{\pi N} = - 0.1 \, {\rm (1a)}, \quad = 0.1 \, {\rm (1b)}, \\ 
&{}& \beta_{1/2}^{\pi N} = \beta_{3/2}^{\pi N} = 1.5 \; {\rm fm^{-1}}, \quad
\lambda_{1/2}^{\pi N} = \lambda_{3/2}^{\pi N} = - 0.03 \, {\rm (2a)}, \quad = 0.03 \, {\rm (2b)}.
\end{eqnarray}
Two more sets of the strength parameters $\lambda_{I}^{\pi N}$  were fitted
to the isospin $I=1/2$ and $I=3/2 $ scattering lengths extracted from experimental data \cite{piN_scL}
\begin{equation}
 a_{\pi N, I=1/2}^{\rm Exp} = 0.26 \; {\rm fm}, \;
  a_{\pi N, I=3/2}^{\rm Exp} = -0.11 \; {\rm fm},
\end{equation}
while the corresponding range parameters $\beta_{I}^{\pi N}$ were chosen by hand. These two sets
of $\pi N$ potential parameters Eq.(\ref{VpiN}) are denoted as "ScL-1" and "ScL-2":
\begin{eqnarray}
 &{}&  \beta_{I=1/2}^{\pi N} = \beta_{3/2}^{\pi N} = 3 \; {\rm fm^{-1}}, \qquad
\lambda_{1/2}^{\pi N} = -0.627 \, {\rm fm}, \,  \lambda_{3/2}^{\pi N} = 0.456 \, {\rm fm} \, {\rm (ScL-1)},\\
 &{}&  \beta_{I=1/2}^{\pi N} = \beta_{3/2}^{\pi N} = 1.5 \; {\rm fm^{-1}}, \quad
\lambda_{1/2}^{\pi N} = -0.046 \, {\rm fm}, \, \lambda_{3/2}^{\pi N} = 0.026 \, {\rm fm}\, {\rm (ScL-2)}.
\end{eqnarray}

\section{Three-body Faddeev-type AGS equations and antikaon-nucleon potentials
with three coupled particle channels}
\label{3AGS3ch_sect}

It was written in Sect.\ref{3AGS2ch_sect} that due to the coupling of the $\bar{K}N$ system with
$\pi \Sigma$ channel, the three-body Faddeev-type AGS equations were solved for
two coupled particle channels: $\bar{K}NN$ and $\pi \Sigma N$. However, antikaon-nucleon
interaction is coupled to the lowest $\pi \Lambda$ channel as well. If we include the third,
$\pi \Lambda$ channel into three-body equations directly, the sum on $\gamma$ index goes
up to $5$:
\begin{equation}
\label{AGS3ch}
 {X}_{ij}^{\alpha \beta}(z) = \delta_{\alpha \beta} {Z}_{ij}^{\alpha} + 
  \sum_{k=1}^3  { \sum_{\gamma=1}^5} {Z}_{ik}^{\alpha} \, {\tau}_{k}^{\alpha \gamma} \,
   {X}_{kj}^{\gamma \beta}
\end{equation}
since now there are five  particle channels
\begin{eqnarray}
\alpha = 1: | \bar{K}_1N_2N_3 \rangle, \qquad \alpha = 2: | \pi_1 \Sigma_2 N_3 \rangle,
\qquad \alpha = 3: | \pi_1 N_2 \Sigma_3 \rangle, \\
{ \alpha = 4: | \pi_1 \Lambda_2 N_3 \rangle,
\qquad \alpha = 5: | \pi_1 N_2 \Lambda_3 \rangle}.
\end{eqnarray}
Antikaon-nucleon potentials with three coupled $\bar{K}N - \pi \Sigma - \pi \Lambda$
channels should be used in the system of integral equations derived from Eq.(\ref{AGS3ch}).
The previously used chirally motivated antikaon-nucleon potential already couples all
three  channels. We refitted parameters of the potential to have strong poles corresponding
to the $\Lambda(1405)$ resonance closer to other chiral models.

The $\pi \Lambda$ channel was taken in our phenomenological  $V^{\rm 1,SIDD}$,
$V^{\rm 2,SIDD}$ potentials into account indirectly through imaginary part of the complex
strength parameter $\lambda^{11}_{I=1}$.
We constructed new versions of the phenomenological potentials directly coupling
all three particle channels $\bar{K}N$, $\pi \Sigma$, and $\pi \Lambda$.
As before, the $V^{\rm 1,SIDD}_{\bar{K}N - \pi \Sigma - \pi \Lambda}$ interaction model
give one pole corresponding to the $\Lambda(1405)$ resonance, while 
the $V^{\rm 2,SIDD}_{\bar{K}N - \pi \Sigma - \pi \Lambda}$ potential lead to two
strong poles for the resonance. All three new antikaon-nucleon $\bar{K}N - \pi \Sigma - \pi \Lambda$
potentials reproduce $1s$ level shift and width of kaonic hydrogen and low energy
cross-sections of $K^- p \to K^- p$ and $K^- p \to M B$ reactions. They also reproduce
two recently measured accurate data on $K^- p \to \pi^0 \Sigma^0$ and $K^- p \to \pi^0 \Lambda$
cross-sections \cite{KpLASTexp}.
In contrast to the previous versions, the new phenomenological
potentials directly reproduce all three threshold branching ratios $\gamma$, $R_c$, and $R_n$.
The same does the chirally motivated model with the new set of parameters. Physical characteristics of
the three new antikaon-nucleon potentials are shown in Table \ref{3chKN_char.tab}.
\begin{table}[ht]
\begin{center}
\begin{tabular}{ccccc}
\hline  \noalign{\smallskip}
  & $V_{\bar{K}N-\pi \Sigma - \pi \Lambda}^{\rm 1,SIDD}$  &  $V_{\bar{K}N-\pi \Sigma - \pi \Lambda}^{\rm 2,SIDD}$  &
    $V_{\bar{K}N-\pi \Sigma - \pi \Lambda}^{\rm Chiral}$  & Exp\\[1mm]
\noalign{\smallskip} \hline \noalign{\smallskip}
 $\Delta E_{1s}$  & $-322.6$ & $-323.5$ & $-311.6$ & $-283\pm 36 \pm 6$ \cite{SIDDHARTA}\\[1mm]
 $\Gamma_{1s}$ & $645.4$ & $633.8$ & $605.8$ & $541 \pm 89 \pm 22$ \cite{SIDDHARTA}\\[1mm]
 $E_1$ & $1429.5 - i \,35.0$ & $ 1430.9 - i\, 41.6$ & $1429.6 - i\, 33.2$ & $-$\\[1mm]
 $E_2$ & $-$ & $ 1380.4 - i \, 79.9$ & $1367.8 - i\, 66.5$ & $-$\\[1mm]
 $\gamma$ & $2.35$ &  $2.36$ & $2.36$ & $2.36 \pm 0.04$ \cite{gammaKp1,gammaKp2}\\[1mm]
 $R_c$         & $0.666$ & $0.664$ & $0.664$ & $0.664 \pm 0.011$ \cite{gammaKp1,gammaKp2}\\[1mm]
 $R_n$        & $0.190$ & $0.189$ & $0.190$ & $0.189 \pm 0.015$ \cite{gammaKp1,gammaKp2}\\[1mm]
 $a_{K^- p}$  & $-0.77 + i\, 0.97$ & $-0.78+ i\, 0.95$ & $-0.75 + i\, 0.90$ &  $-$\\[1mm]
 \noalign{\smallskip} \hline
\end{tabular}
\caption{Physical characteristics of the three $\bar{K}N -  \pi \Sigma - \pi \Lambda$ potentials:
$1s$ level shift $\Delta E_{1s}$  and width  $\Gamma_{1s}$, strong poles $E_1$ and $E_2$, 
threshold branching ratios $\gamma$, $R_c$, $R_n$. The $K^- p$ scattering length
$a_{K^- p}$ was calculated with physical masses in all channels and Coulomb interaction in
$K^- p$ channel. Experimental data are also shown.
\label{3chKN_char.tab}
}
\end{center}
\end{table}

\section{Results obtained with two coupled $\bar{K}NN - \pi \Sigma N$ channels}
\label{results2ch_sect}

The results of the three-body $\bar{K}NN - \pi \Sigma N$ calculations with the new
$\Sigma N - \Lambda N$ potentials are shown in Table \ref{SigmaN_dep.tab}. The three models
of the antikaon-nucleon interaction were used: one- $V_{\bar{K}N}^{\rm 1,SIDD}$ and
two-pole $V_{\bar{K}N}^{\rm 2,SIDD}$ phenomenological potentials and the chirally
motivated potential $V_{\bar{K}N}^{\rm Chiral}$.  The $\pi N$ interaction was switched off,
as in our previous calculations. The first line contains our previous results \cite{Kd_Kpp_last}
obtained with the older $\Sigma N - \Lambda N$ interaction model, denoted as
$V_{\Sigma N-\Lambda N}^{\rm Prev}$. The binding energies of the $K^- pp$ quasi-bound
state $B_{K^-pp}$ and its widths  $\Gamma_{K^-pp}$ were calculated with the four new sets 
of parameters of the hyperon-nucleon interaction model: spin-dependent ("SDep") or
spin-independent ("SInd") ones with ("ScL") or without ("noScL") fit to hyperon-nucleon
scattering lengths from the "advanced" potential \cite{HyperN_poten}.
\begin{table}[hb]
\begin{center}
\begin{tabular}{ccccccc}
\hline  \noalign{\smallskip}
  & \multicolumn{2}{c}{$V_{\bar{K}N - \pi \Sigma}^{\rm 1,SIDD}$} 
    & \multicolumn{2}{c}{$V_{\bar{K}N - \pi \Sigma}^{\rm 2,SIDD}$}  
     & \multicolumn{2}{c}{$V_{\bar{K}N - \pi \Sigma (- \pi \Lambda)}^{\rm Chiral}$}  
     \\[1mm]
  & $B_{K^-pp}$ \; &  $\Gamma_{K^-pp}$ \;  & $B_{K^-pp}$ \;  & $\Gamma_{K^-pp}$ \;  
        & $B_{K^-pp}$ \;  & $\Gamma_{K^-pp}$  \\[1mm]
\noalign{\smallskip} \hline \noalign{\smallskip}
   $V_{\Sigma N-\Lambda N}^{\rm Prev}$ & $52.2$  & $67.1$ & $46.6$ & $51.2$ & $29.4$ & $46.4$   \\[1mm]
\noalign{\smallskip} \hline \noalign{\smallskip}
   $V_{\Sigma N-\Lambda N}^{\rm noScL,SInd}$ & $63.5$  & $96.5$ & $52.2$ & $61.0$ & $28.3$ & $49.5$   \\[2mm]
$V_{\Sigma N-\Lambda N}^{\rm noScL,SDep}$ & $52.5$  & $67.0$ & $46.1$ & $49.8$ & $29.6$ & $46.8$   \\[2mm]
   $V_{\Sigma N-\Lambda N}^{\rm ScL,SInd}$ & $38.1$  & $48.9$ & $35.4$ & $41.1$ & $29.5$ & $39.3$   \\[2mm]
$V_{\Sigma N-\Lambda N}^{\rm ScL,SDep}$ & $34.3$  & $65.6$ & $34.7$ & $53.4$ & $27.9$ & $42.8$
  \\[2mm]
 \noalign{\smallskip} \hline
\end{tabular}
\caption{Dependence of the binding energy $B_{K^-pp}$ (MeV) and width $\Gamma_{K^-pp}$ (MeV)
of the quasi-bound state in the {$K^- pp$ system} on $\Sigma N - \Lambda N$ interaction models. 
Previous results evaluated with $V_{\Sigma N-\Lambda N}^{\rm Prev}$ from \cite{Kd_Kpp_last} are
also shown. All observables were obtained with switched off $\pi N$ potential. Three-body calculations 
were performed with two coupled $\bar{K}NN - \pi \Sigma N$ channels.
\label{SigmaN_dep.tab} 
}
\end{center}
\end{table}

It is seen from Table \ref{SigmaN_dep.tab} that the dependence of the three-body results on
the hyperon-nucleon potential can be quite strong, especially when the phenomenological
$\bar{K}N - \pi \Sigma$ potentials are used. Influence of the $V_{\Sigma N-\Lambda N}$ being used
together with the chirally motivated $\bar{K}N - \pi \Sigma (- \pi \Lambda)$ potential is much weaker.
Surprisingly, not only width of the $K^- pp$ state is influenced by the $\Sigma N - \Lambda N$
interaction, but its binding energy also can be changed sufficiently. Probably it is caused by the fact,
that some of the four hyperon-nucleon potentials are characterized by very strong attraction in
the $\Sigma N /\Lambda N$ pairs.

The most reliable hyperon-nucleon model among the four is the spin-dependent potential
$V_{\Sigma N-\Lambda N}^{\rm ScL,SDep}$ with parameters fitted to experimental cross-sections
and scattering lengths. The  resulting binding energies calculated with all three antikaon-nucleon
potentials turned out to be smaller than the initial ones calculated with the previous 
$V_{\Sigma N - \Lambda N}^{\rm Prev}$. The difference is much larger when the phenomenological
models of $V_{\bar{K}N - \pi \Sigma}$ potential are used. The widths are changed slightly.

Dependence of the results of the three-body calculations with two coupled $\bar{K}NN - \pi \Sigma N$
channels on pion-nucleon interaction is shown in Table \ref{piN_dep.tab}. Binding energy $B_{K^-pp}$
and width  $\Gamma_{K^-pp}$ of the $K^- pp$ quasi-bound state obtained using six new $\pi N$ potentials 
are compared with the previously calculated results with switched off $\pi N$ potential, shown at the first line. 
All old and new results were evaluated using the best hyperon-nucleon 
$V_{\Sigma N-\Lambda N}^{\rm ScL,SDep}$ potential. The four versions of the $\pi N$ potential with
arbitrary parameters $V_{\pi N}^{\rm (1a)}$, $V_{\pi N}^{\rm (1b)}$, $V_{\pi N}^{\rm (2a)}$, and
$V_{\pi N}^{\rm (2b)}$ lead to negligible changes in the binding energy obtained with every of the
three $\bar{K}N$ potentials. The dependence of the three-body widths on the $\pi N$ interaction model
is more visible, especially when one of the phenomenological antikaon-nucleon potentials is used. 
Four $\pi N$ interaction models change the width by few MeV. 
Changes in the three-body results are stronger when the $\pi N$ models fitted to scattering
lengths $V_{\pi N}^{\rm ScL-1}$ and $V_{\pi N}^{\rm ScL-2}$ are used. However, the dependence of
the $K^- pp$ quasi-bound state binding energy and width  on the $\pi N$ interaction models is 
quite weak, it is much weaker than on the hyperon-nucleon interaction.

Comparing the first line of Table \ref{SigmaN_dep.tab}, containing the previous results
from \cite{Kd_Kpp_last},  with the two last lines of  Table \ref{piN_dep.tab}, we see the cumulative
effect of the ''less important''  pion-nucleon and hyperon-nucleon interactions. It is seen that
the results of the three-body calculations with two coupled $\bar{K}NN - \pi \Sigma N$
channels evaluated with the chirally motivated model of antikaon-nucleon interaction are
influenced by the $\Sigma N - \Lambda N$ and $\pi N$ interactions slightly. These interactions
in the lower channel are much more important together with the phenomenological
models. Using of the accurate $V_{\Sigma N - \Lambda N}$ and $V_{\pi N}$ potentials lead
to much smaller binding energies $B_{K^-pp}$ and larger widths $\Gamma_{K^-pp}$ of
the $K^- pp$ system calculated with $V_{\bar{K}N}^{\rm 1,SIDD}$ and $V_{\bar{K}N}^{\rm 2,SIDD}$.

\begin{table}[ht]
\begin{center}
\begin{tabular}{ccccccc}
\hline  \noalign{\smallskip}
  & \multicolumn{2}{c}{$V_{\bar{K}N - \pi \Sigma}^{\rm 1,SIDD}$} 
    & \multicolumn{2}{c}{$V_{\bar{K}N - \pi \Sigma}^{\rm 2,SIDD}$}  
     & \multicolumn{2}{c}{$V_{\bar{K}N - \pi \Sigma (- \pi \Lambda)}^{\rm Chiral}$}  
     \\[1mm]
  & $B_{K^-pp}$ \; &  $\Gamma_{K^-pp}$ \;  & $B_{K^-pp}$ \;  & $\Gamma_{K^-pp}$ \;  
        & $B_{K^-pp}$ \;  & $\Gamma_{K^-pp}$  \\[1mm]
\noalign{\smallskip} \hline \noalign{\smallskip}
   $V_{\pi N}^{\rm off}$ & $34.3$  & $65.6$ & $34.7$ & $53.4$ & $27.9$ & $42.8$   \\[1mm]
\noalign{\smallskip} \hline \noalign{\smallskip}
   $V_{\pi N}^{\rm (1a)}$ & $34.6$  & $66.4$ & $34.9$ & $54.2$ & $27.7$ & $43.3$   \\[2mm]
$V_{\pi N}^{\rm (1b)}$ & $34.1$  & $64.8$ & $34.4$ & $52.7$ & $28.0$ & $42.4$   \\[2mm]
   $V_{\pi N}^{\rm (2a)}$ & $34.6$  & $69.5$ & $35.1$ & $56.6$ & $27.1$ & $43.9$   \\[2mm]
$V_{\pi N}^{\rm (2b)}$ & $33.9$  & $62.3$ & $34.1$ & $50.8$ & $28.4$ & $41.6$    \\[2mm]
\noalign{\smallskip} \hline \noalign{\smallskip}
   $V_{\pi N}^{\rm ScL-1}$ & $37.6$  & $71.4$ & $37.7$ & $58.2$ & $27.2$ & $46.6$   \\[2mm]
$V_{\pi N}^{\rm ScL-2}$ & $35.7$  & $70.5$ & $35.9$ & $57.4$ & $27.0$ & $44.6$    \\
 \noalign{\smallskip} \hline
\end{tabular}
\caption{Dependence of the binding energy $B_{K^- pp}$ (MeV) and width $\Gamma_{K^- pp}$ (MeV)
of the quasi-bound state in the {$K^- pp$} system on $\pi N$ interaction models.  All observables
were calculated with  $V_{\Sigma N-\Lambda N}^{\rm ScL,SDep}$ potential, including the previous
ones from Table \ref{SigmaN_dep.tab} with switched off pion-nucleon interaction $V_{\pi N}^{\rm off}$.
The three-body calculations were performed with two coupled $\bar{K}NN - \pi \Sigma N$ channels.
\label{piN_dep.tab}
}
\end{center}
\end{table}

\section{Results obtained with three coupled $\bar{K}NN - \pi \Sigma N - \pi \Lambda N$ channels}
\label{results3ch_sect}

The results of the direct inclusion of the lowest $\pi \Sigma N$ channel in the 
$\bar{K}NN - \pi \Sigma N - \pi \Lambda N$ system are shown in Table \ref{3ch_res.tab}. 
The calculations were performed with the newly constructed phenomenological potentials
directly coupling all three particle channels $V^{\rm 1,SIDD}_{\bar{K}N - \pi \Sigma - \pi \Lambda}$
and $V^{\rm 2,SIDD}_{\bar{K}N - \pi \Sigma - \pi \Lambda}$  together with the chirally motivated potential
$V^{\rm Chiral}_{\bar{K}N - \pi \Sigma - \pi \Lambda}$ with refitted parameters. The new potentials are
described in Section \ref{3AGS3ch_sect}. The most accurate  $V_{\Sigma N-\Lambda N}^{\rm ScL,SDep}$
potential and the two versions of the $\pi N$ interaction model  $V_{\pi N}^{\rm ScL-1}$,
$V_{\pi N}^{\rm ScL-2}$ were used in the calculations. 

It is seen that the two-pole phenomenological and chiral potentials give approximately equal
widths of the quasi-bound $K^- pp$ state, while the width obtained with the one-pole
phenomenological potential is smaller. As before, the smallest $K^- pp$ binding energy
is the result of using the chirally motivated potential, the difference between the one-pole
and two-pole phenomenological results is now larger than before. Comparing the results of
the three-body calculations with two coupled $\bar{K}NN - \pi \Sigma N$ in Table \ref{piN_dep.tab} 
and three coupled $\bar{K}NN - \pi \Sigma N - \pi \Lambda N$ channels in Table \ref{3ch_res.tab}
we see large differences between the observables. However, it is not clear, whether the difference
is caused by the three-channel structure or by using different $\bar{K}N$ potentials leading to different
two-body observables.

\begin{table}[ht]
\begin{center}
\begin{tabular}{ccccccc}
\hline  \noalign{\smallskip}
  & \multicolumn{2}{c}{$V_{\bar{K}N - \pi \Sigma - \pi \Lambda}^{\rm 1,SIDD}$} 
    & \multicolumn{2}{c}{$V_{\bar{K}N - \pi \Sigma - \pi \Lambda}^{\rm 2,SIDD}$}  
     & \multicolumn{2}{c}{$V_{\bar{K}N - \pi \Sigma - \pi \Lambda}^{\rm Chiral}$}  
     \\[1mm]
  & $B_{K^-pp}$ \; &  $\Gamma_{K^-pp}$ \;  & $B_{K^-pp}$ \;  & $\Gamma_{K^-pp}$ \;  
        & $B_{K^-pp}$ \;  & $\Gamma_{K^-pp}$  \\[1mm]
\noalign{\smallskip} \hline \noalign{\smallskip}
   $V_{\pi N}^{\rm ScL-1} $ & $34.5$  & $52.0$ & $42.9$ & $60.4$ & $26.8$ & $59.6$   \\[1mm]
   $ V_{\pi N}^{\rm ScL-2} $ & $34.3$  & $52.1$ & $40.2$ & $57.7$ & $27.2$ & $56.3$   \\[1mm]
 \noalign{\smallskip} \hline
\end{tabular}
\caption{Results of the three-body AGS calculations with three coupled
$\bar{K}NN - \pi \Sigma N - \pi \Sigma \Lambda $ channels. Binding energy $B_{K^-pp}$ (MeV) and
width $\Gamma_{K^-pp}$ (MeV) of the quasi-bound state in the {$K^- pp$} system were evaluated
using $V_{\Sigma N-\Lambda N}^{\rm ScL,SDep}$ potential and two pion-nucleon 
interaction models $V_{\pi N}^{\rm ScL-1}$ and $V_{\pi N}^{\rm ScL-2}$.
\label{3ch_res.tab}
}
\end{center}
\end{table}

\section{Summary}
\label{Summary_sect}

Fine-tuning of the dynamically exact calculations of the binding energy and width of the $K^- pp$
quasi-bound state was performed. Influence of the "less important" interactions in the lower
$\pi \Sigma N$ channel was studied in three-body Faddeev-type calculations with two coupled
$\bar{K}NN - \pi \Sigma N$ channels. It was found that the $\Sigma N - \Lambda N$ interaction
model can change the three-body result quite strongly, while the dependence of the $B_{K^-pp}$ and 
$\Gamma_{K^-pp}$  on the $\pi N$ potential is weaker.  The new antikaon-nucleon potentials
coupling three $\bar{K}N - \pi \Sigma - \pi \Lambda$ channels were constructed. The corresponding
three-body calculations of the $K^- pp$ quasi-bound state with three coupled
$\bar{K}NN - \pi \Sigma N - \pi \Lambda N$ channels were performed. The three-body results
evaluated with three coupled channels lead to different changes in binding energy and/or
width in comparison with the calculations with two coupled channels depending on the particular
model of antikaon-nucleon interaction. However, the reason for the differences:  whether it is new
potentials or the third coupled channel, - should be additionally studied.


\end{document}